\documentclass[%
 aip,
 amsmath,amssymb,
 reprint,%
]{revtex4-1}

\usepackage{graphicx}
\usepackage{dcolumn}
\usepackage{bm}

\usepackage[utf8]{inputenc}
\usepackage[T1]{fontenc}
\usepackage{mathptmx}
\usepackage{etoolbox}
\usepackage{float}
\usepackage{bm}

\makeatletter
\def\@email#1#2{%
 \endgroup
 \patchcmd{\titleblock@produce}
  {\frontmatter@RRAPformat}
  {\frontmatter@RRAPformat{\produce@RRAP{*#1\href{mailto:#2}{#2}}}\frontmatter@RRAPformat}
  {}{}
}%
\makeatother
\begin{document}

\preprint{AIP/123-QED}

\title[Modeling thermocapillary microgear rotation and transfer to translational particle propulsion]{Modeling thermocapillary microgear rotation and transfer to translational particle propulsion}
\author{Tillmann Carl}

 \affiliation{\textsuperscript{\normalfont\ 1)}Rheinland-Pfälzische Technische Universität Kaiserslautern-Landau, 67663, Kaiserslautern, Germany}
\author{Clarissa Schönecker}
\affiliation{\textsuperscript{\normalfont\ 1)}Rheinland-Pfälzische Technische Universität Kaiserslautern-Landau, 67663, Kaiserslautern, Germany}

\date{\today}

\begin{abstract}
In this study, we investigate the thermocapillary rotation of microgears at fluid interfaces and extend the concept of geometric asymmetry to the translational propulsion of micron-sized particles. We introduce a transient numerical model that couples the Navier-Stokes equations with heat transfer, displaying particle motion through a moving mesh interface. The model incorporates absorbed light illumination as a heat source and predicts both rotational and translational speeds of particles. Our simulations explore the influence of microgear design geometry and determine the scale at which thermocapillary Marangoni motion could serve as a viable propulsion method. A clear correlation between Reynolds number and propulsion efficiency can be recognized. To transfer the asymmetry-based propulsion principle from rotational to directed translational motion, various particle geometries are considered. The exploration of breaking geometric symmetry for translational propulsion is mostly ignored in the existing literature, thus warranting further discussion. Therefore, we analyse expected translational speeds in comparison to corresponding microgears to provide insights into this promising propulsion method.
 \end{abstract}

\maketitle

\section{\label{sec:introduction}Introduction}
The Marangoni effect, induced by surface tension gradients on a fluid interface, offers a mechanism for fluid flow manipulation. Various strategies exploit this effect to propel small objects swimming at fluid interfaces. These strategies include solutocapillarity, which manipulates surface tension via chemical or surfactant addition, and thermocapillarity, which leverages temperature-induced surface tension changes. While solutocapillarity relies on consumable "fuel" and can suffer from saturation effects, thermocapillarity offers a durable, fuel-free alternative.

At small length scales dominated by viscous effects, Marangoni-induced interfacial motion has garnered significant interest for propelling micron-sized particles\cite{maggi2015micromotors,sanchez2015chemically,wang2013small}, driving advancements in self-propelled nano- and micromachines. In this context and beyond, thermocapillary but also thermophoretic and bacterial-driven gear-shaped micromotors have been developed\cite{maggi2015micromotors,yang2014self,angelani2009self}, serving as promising tools for future applications in microfluidics. In the context of thermocapillarity, a common scenario involves a micron-sized, gear-shaped particle with a specific number of teeth, swimming on a liquid-gas interface. Upon heating, such as through illumination from above, interfacial thermocapillary forces rise and initiate a rotational motion of the gear. In the first part of this study, we focus on characterizing the thermocapillary rotation of such microgears by means of numerical simulations, exploring the influence of gear geometry by varying its number of teeth and the influence of the size scale on achievable rotational speeds.

To initiate rotational or translational motion in particles by thermocapillary means, a nonuniform surrounding temperature distribution is required, creating nonuniform stresses. This can for example be achieved through asymmetric coatings that absorb heat to different degrees, such as Janus particles\cite{PhysRevLett.125.098001}. Alternatively, symmetry-breaking in geometry can induce asymmetric temperature gradients, as demonstrated by microgears. This approach is well-known\cite{wang2013small,PhysRevLett.125.098001}. However, while translational propulsion via asymmetric coatings\cite{PhysRevLett.125.098001}  or chemical release\cite{yang2014self,PhysRevLett.125.098001,masoud2014reciprocal,sur2019translational,crowdy2021viscous,kang2020forward} has received considerable attention in the literature, investigations into geometry-induced symmetry breaking for translational thermocapillary propulsion have not yet been described.

To address this gap, in the second part of our study, we introduce and explore thermocapillary-driven translational motion of illuminated asymmetric particles swimming at fluid interfaces. In this context, various geometric shapes are investigated, comparing their achievable propulsion speeds to rotating microgears under similar illumination conditions. We employ the same numerical model as used for rotational motion of thermocapillary microgears, with necessary adjustments to account for translational dynamics.

\section{\label{sec:mandm}Mathematical framework and modeling approach\protect\\}
The model presented  describes a scenario in which an asymmetric microgear is positioned on a fluid (liquid-gas) interface. Illumination directed from above induces heat absorption by the particle. Consequently, an asymmetric temperature distribution forms around the particle, giving rise to thermocapillary forces that induce a rotational motion of the particle. The model allows the calculation of the rotational speed under certain conditions, including illumination power and absorption coefficient, general material properties of the solid particle and the liquid as well as the particle geometry. Analogously, it predicts the propulsion speed of transitional motion.

\begin{figure}[H]
    \centerline{\includegraphics[scale=0.75]{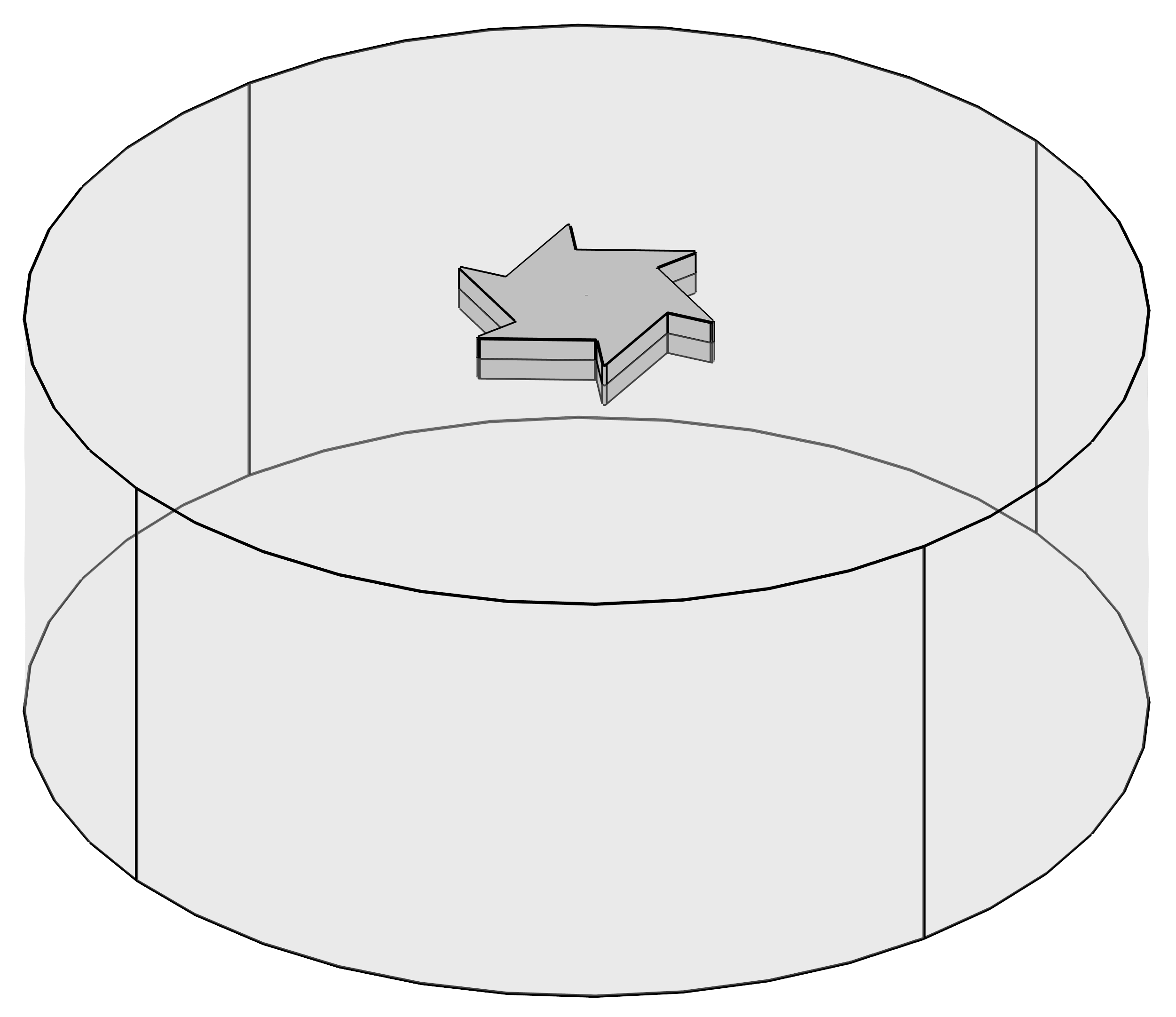}}
         \caption{Simulation domain for a microgear rotation scenario. The particle sits on the surface of the cylindrical fluid domain, half immersed in the liquid.}
  \label{fig:domain}   
\end{figure}

For the scenario of a rotating microgear, the simulation setup comprises a cylindrical fluid domain with a solid particle body situated at the center of the top surface, half immersed in the fluid (see figure\,\ref{fig:domain}). The fluid domain has a height of 26\,µm, which is ten times the height of the particle, and a diameter of 64\,µm, which is four times the diameter of the particle.
\noindent
The dimensions of the fluid domain are chosen to be as small as possible, without affecting physics around the particle. As criterion, the stationary temperature distribution and hence the Marangoni forces are considered. Test calculations indicate that this is the case for the abovementioned dimensions.

The flow in the fluid domain is characterized by the transient Navier-Stokes equation
\begin{equation}
	\rho  \frac{\partial\vec{u} }{\partial t}+\rho(\vec{u}\cdot\nabla)\vec{u}=-\nabla p+\mu \Delta \vec{u}+\rho\vec{g},
	\label{eq:navierstokes}
\end{equation}
and the incompressible equation for mass conservation
\begin{equation}
	\nabla\cdot \vec{u} =0,
	\label{eq:konti}
\end{equation}
with the density $\rho$, the fluid velocity $\vec{u}=(u,v,w)^T$, the pressure $p$, the dynamic viscosity $\mu$, and the gravitational force $\vec{g}$. The assumption of incompressibility neglects the variation in density caused by changes in temperature. However, for very small  temperature changes, as is the case in this work, this assumption is reasonable. On micron scale, where viscous forces dominate, equation\,\eqref{eq:navierstokes} could in principle be simplified to the Stokes equation
\begin{equation}
	\rho  \frac{\partial\vec{u} }{\partial t}=-\nabla p+\mu \Delta \vec{u}+\rho\vec{g}.
	\label{eq:stokes}
\end{equation}
As will be seen in the course of this work, the Stokes assumption will break down for certain larger particles under high illumination. Therefore, the Navier-Stokes equation is used throughout for the sake of consistency.  To maintain the effect of an infinite domain, the outer boundary walls of the fluid domain are modelled as shear-free. This ensures that the fluid flow near the walls is not artificially slowed down. For the walls at the particle interface, the no-slip condition applies.

\begin{figure}[H]
    \centerline{\includegraphics[scale=0.57]{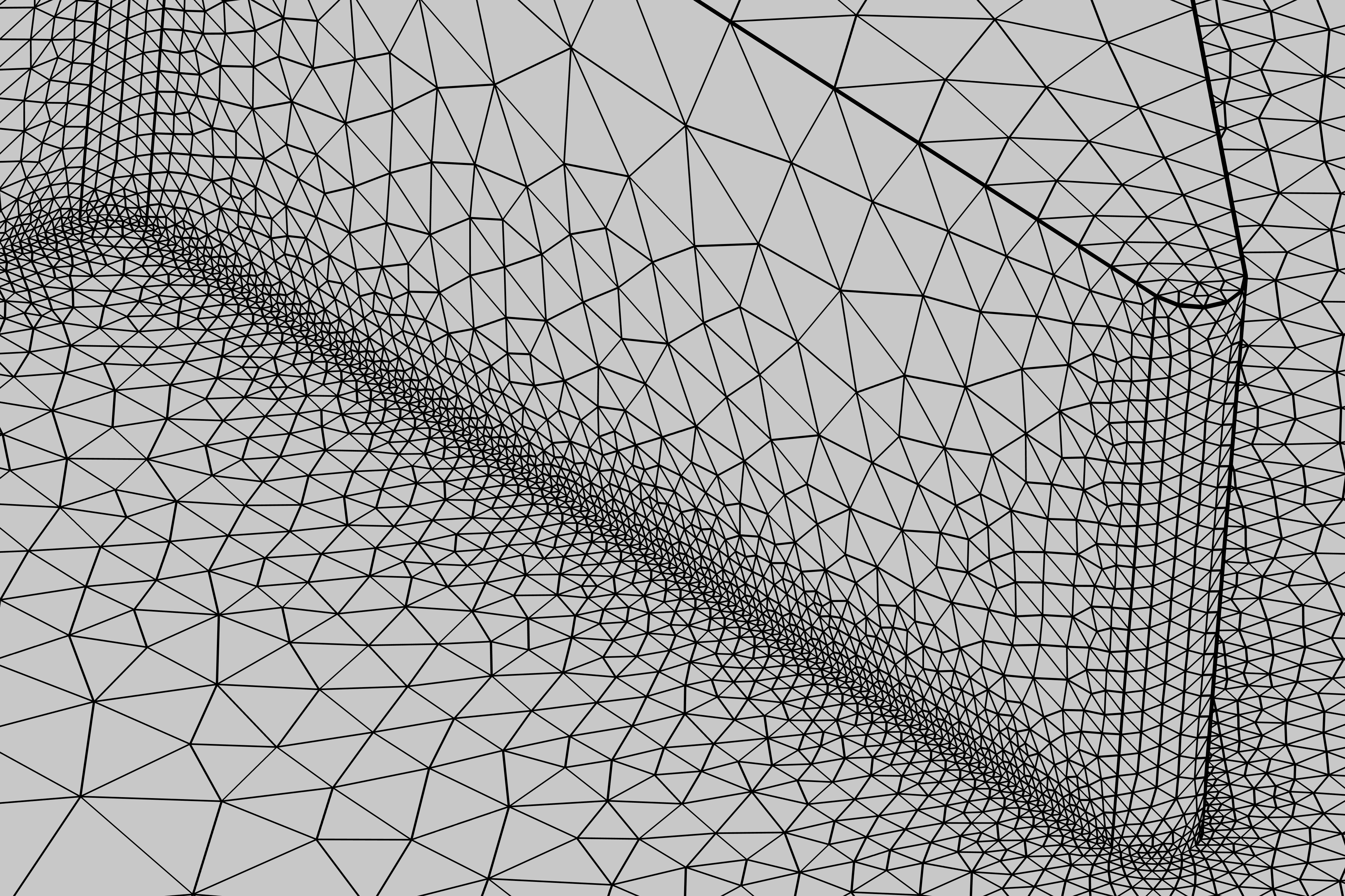}}
    \caption{The fine mesh around the immersed particle wall leads to an element count of around 2.7 million cells for microgear measurements as in section \ref{sec:ValRes}. The resulting simulation time for these calculations amounts to a two-digit number of hours with an 8-core CPU (AMD Ryzen 7) and 64\,GB of RAM.}
    \label{fig:mesh}
\end{figure}

In both domains, the solid particle and the fluid, transient heat transfer is occurring
\begin{equation}
\rho c_{p}\frac{\partial{T}}{\partial{t}}+\rho c_{p} \vec{u} \nabla T +\nabla \cdot (-\lambda \nabla T)=Q.
    \label{eq:heattransfer}
\end{equation}
 Here, $\vec{u}$ represents the fluid velocity in the fluid domain. In the solid domain, it is replaced by $\vec{u}_s$, representing  the velocity field defined by the moving mesh node of the moving solid particle, considering a corresponding motion of the heat source. The parameters $\lambda$ and $c_{p}$ are the thermal conductivity and the heat capacity of the corresponding domain. The heat source $Q$ represents the absorbed illumination power of the particle and is set as a boundary condition on the top particle surface, being zero everywhere else. A detailed explanation on illumination and the absorption coefficient is provided in section \ref{sec:ValRes}. At the outer boundary walls, a fixed temperature of 293.15\,K is set. The temperature difference between the outer wall and the heated particle leads to thermocapillary forces in the fluid. These resulting thermocapillary forces act on the top surface of the fluid domain, which represents the fluid interface (liquid/gas). There, continuity of velocity and shear stress holds

\begin{subequations}
\label{eq:shearcont}
\begin{equation}
\mu_{l}\frac{\partial u_{l} }{\partial x}+f_{x}=\mu_{g}\frac{\partial u_{g} }{\partial x},
\label{eq:shearcontx}
\end{equation}
\begin{equation}
\mu_{l}\frac{\partial v_{l} }{\partial y}+f_{y}=\mu_{g}\frac{\partial v_{g} }{\partial y}.
\label{eq:shearconty}
\end{equation}
\end{subequations}\noindent
Since the dynamic viscosity $\mu_{g}$ is small compared to that of the liquid, the term on the right side can be neglected in equations \eqref{eq:shearcontx} and \eqref{eq:shearconty}, setting the right side to zero. This assumes a drag-free air interface. The thermocapillary forces $\vec{f}$ are defined as
\begin{equation}
\vec{f}=
\begin{pmatrix}
    f_{x}\\
    f_{y}\\
    f_{z}
\end{pmatrix}
=
\begin{pmatrix}
\frac{\partial \sigma}{\partial x} \\
\frac{\partial \sigma}{\partial y}\\
0
\end{pmatrix}
=
\begin{pmatrix}
\frac{\partial \sigma}{\partial T} \frac{\partial T}{\partial x} \\
\frac{\partial \sigma}{\partial T} \frac{\partial T}{\partial y} \\
0
\end{pmatrix}.
	\label{eq:boundaryforce}
\end{equation}
Equations \eqref{eq:shearcont} and \eqref{eq:boundaryforce} can be implemented via a boundary stress feature on the top of the fluid domain that requires only the specification of $\vec{f}$
\begin{equation}
	(-p\vec{I}+\mu (\nabla \vec{u}+(\nabla \vec{u})^T))\vec{n} =\vec{f},
	\label{eq:boundary}
\end{equation}
 with the identity tensor $\vec{I}$ and the normal vector $\vec{n}$. The term $\mu (\nabla \vec{u}+(\nabla \vec{u})^T))$ represents the viscous stress contribution due to (Newtonian) fluid motion. In the next step, the interfacial forces cause a fluid flow, which in turn creates forces $\vec{F}$ on the immersed solid particle wall $A_{c}$. These forces acting on the particle result, due to the asymmetric microgear shape, in a net torque, and are therefore initiating a rotational motion of the gear. 
 Before further describing the implementation of corresponding equations, it has to be mentioned that here, the right meshing strategy is crucial. In order to resolve the temperature gradients at the contact line and the forces acting on the particle, it is necessary to have a highly refined mesh around the contact line, as shown in figure\,\ref{fig:mesh}, and around the immersed particle wall. Although significantly increasing computation time, to obtain mesh-independent results, this is essential. To minimize the mesh influence for comparisons of different geometries, the amount of cells per micrometer at the particle wall is kept constant for all geometries.
\noindent
At the finely meshed immersed particle walls, the forces $\vec{F}$ that the fluid exerts on the particle are evaluated and used to determine the rotational speed $\omega$ of the particle, defined by the equation for rotational dynamics
\begin{equation}
	\frac{\partial \omega}{\partial t}=\frac{M_{\mathrm{tot}}}{J},
	\label{dynamicrot}
 \end{equation}
with the particle's moment of inertia $J$
\begin{equation}
J= \int_{V} |\vec{r}|^2\rho\, \mathrm{d}V,
\label{momofinertia}
\end{equation}
and the total torque $M_{\mathrm{tot}}$ acting on the particle
\begin{equation}
M_{\mathrm{tot}}=\int_{A_{c}} (\vec{r}\times \vec{F})\vec{n} \,\,\mathrm{d}A_{c},
    \label{mom}
\end{equation}
with the local particle radius $\vec{r}$ and $\times$ representing the cross product. To simulate actual motion, we use a moving mesh interface. The calculated rotational speed $\omega$ is taken as input for the particle's mesh rotation. To prevent distortion of the mesh cells around the particle, a defined mesh slip is used at the outer boundary domain, allowing the mesh to move (only) in the current tangential direction. This lets the mesh of the fluid domain rotate with the particle.

The commercial software COMSOL Multiphysics\textsuperscript{\textregistered} is utilized to solve the set of equations.

\section{\label{sec:ValRes}Model validation\protect\\}
\begin{figure}[H]
    \centerline{\includegraphics[scale=0.6]{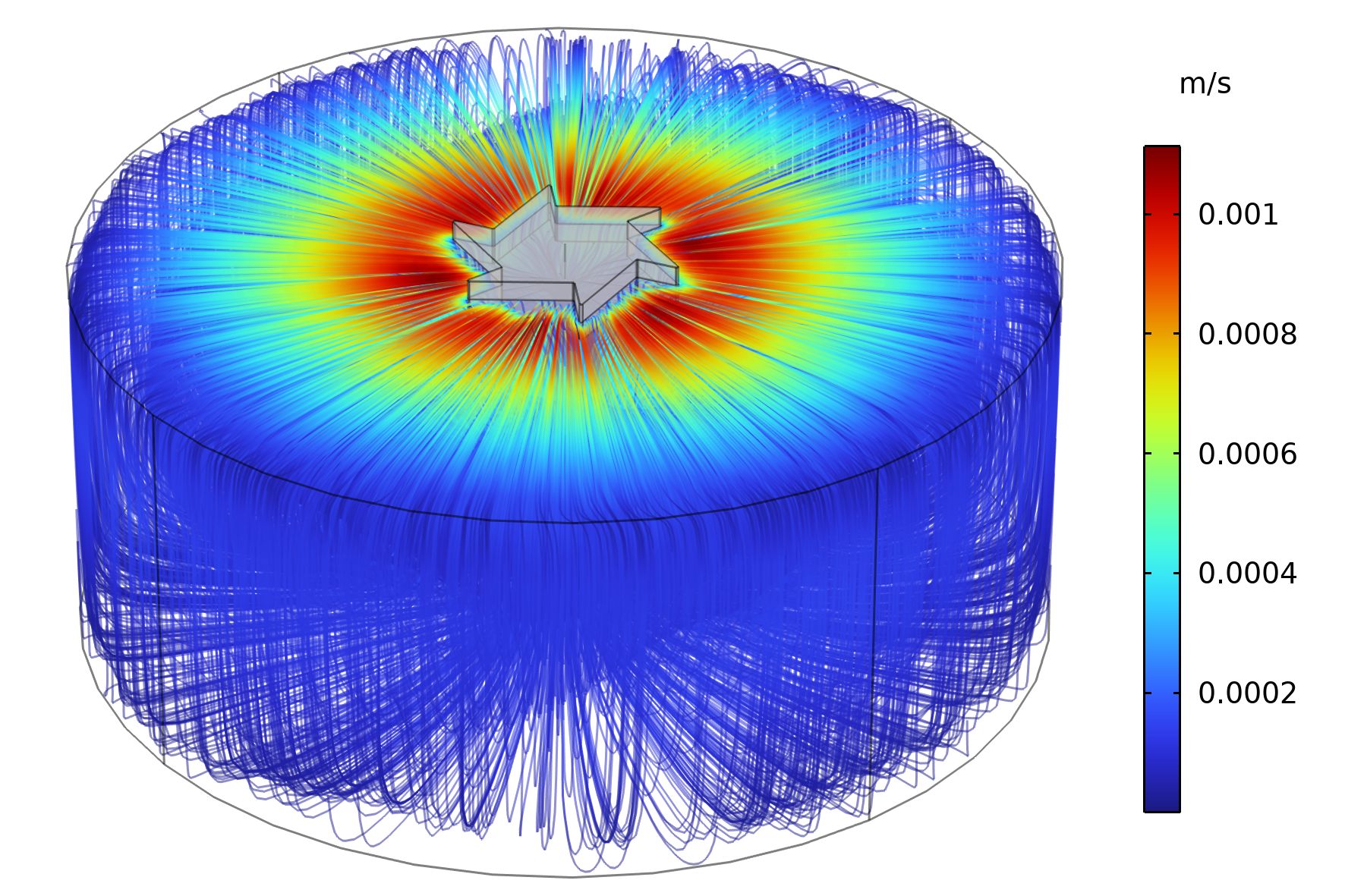}}
    \caption{Marangoni-induced flow field corresponding to an absorbed illumination of 0.8\,µW and a particle rotational speed of about 134\,rpm. }
    \label{fig:flow}
\end{figure}

For validation purpose, the results are compared to the experimental data by Maggi et al.\cite{maggi2015micromotors}, illuminating a  six-toothed microgear of 2.6\,µm in thickness and 16\,µm in diameter with various illumination powers and measuring the corresponding rotational speed. Therefore, the materials used in the experiment are characterized by the following properties (see table\,\ref{props}). For numerical reasons, the inner and outer edges of the gear are rounded with a radius of 0.1\,µm.
\renewcommand{\arraystretch}{1.5}
\begin{table} [H]
\normalsize

\caption{\label{tab:table1} Assumed material properties characterizing the solvent  ${\textrm{N-Methyl-2-pyrrolidone}}$ (liquid) and hardened SU-8 photoresist (solid particle).}
\begin{ruledtabular}
\begin{tabular}{lcr}
\label{props}

Property&Liquid&Solid\\
\hline

Density $\rho$ /$\frac{g}{cm^3}$ & 1.033 & 1.19\\
Thermal conductivity $\lambda$ /$\frac{W}{m \cdot K}$ & 0.187 & 0.2\\
Heat capacity $c_{p}$ /$\frac{kJ}{kg \cdot K}$ & 1.77& 1.6\\
Dynamic viscosity $\mu$ /($mPa \cdot s$) & 1.8 & -\\
Temperature derivative\\ of the surface tension  $\frac{d\sigma}{dT}$ /$\frac{mN}{m\cdot K}$ & -0.11 & -

\end{tabular}
\end{ruledtabular}
\end{table}
In an experimental setup, when illuminating a particle with a light source of the power P, the incident illumination $\mathrm{P_{inc}}$ on the particle surface is defined by the ratio of the particle surface $A_{\mathrm{P}}$ and the illumination area of the light source $A_{\mathrm{l}}$
\begin{equation}
\mathrm{P_{inc}}=\mathrm{P}\frac{A_{\mathrm{P}}}{A_{\mathrm{l}}}.
    \label{arearatio}
\end{equation}
The absorbed illumination power $\mathrm{P_{abs}}$ of the particle is thus defined as
\begin{equation}
\mathrm{P_{abs}}=c_{\mathrm{abs}}\mathrm{P_{inc}},
    \label{abspower}
\end{equation}
with the absorption coefficient $c_{\mathrm{abs}}$. Due to differences in the carbon coating of the particles, varying light absorption coefficients occur. Therefore, various representative absorption coefficient are implemented in the model.
The rotational speeds of the microgears are calculated over increasing illumination power. Figure\,\ref{fig:flow} illustrates the corresponding flow field for an exemplary scenario. As a consequence of the thermocapillary forces, the fluid flows from the heated particle wall (low surface tension) away to the outer  boundary wall (low temperature, high surface tension), propelling the microgear.\\
\noindent

\begin{figure}[H]
    \centerline{\includegraphics[scale=0.6]{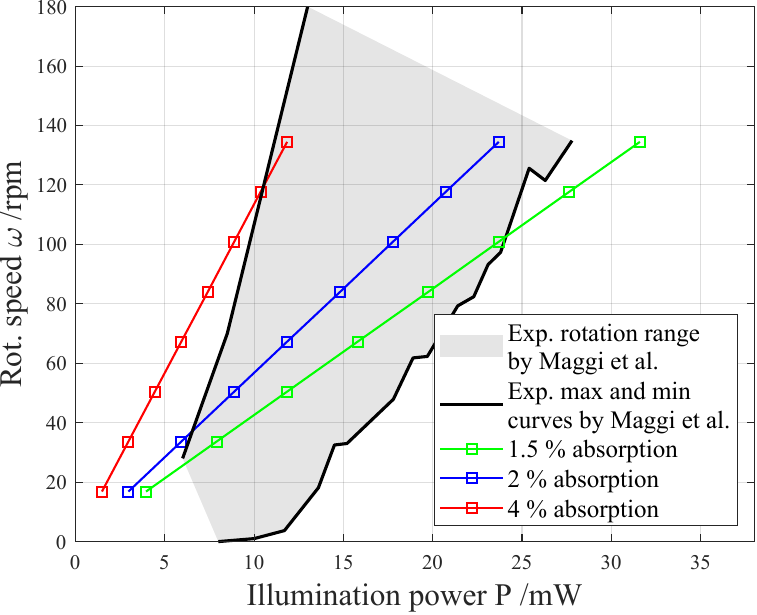}}
    \caption{Comparison of numerical results with experimental data by Maggi et al\cite{maggi2015micromotors}. Displayed are rotational speeds for several absorption coefficients over the illumination power. Experimentally, several curves were measured, all linearly increasing with the illumination power and ranging in between the black curves.}
    \label{fig:Validation}
\end{figure}\noindent
The simulated rotational speeds are of the same order of magnitude as in the experiments (see figure\,\ref{fig:Validation}). In both cases, a linearity between the rotational speed of the microgear and the illumination power is visible. A deviation can be recognized at low illumination powers, where linearity between rotational speed and illumination power is present in the numerical case, unlike in the experiment. This might have several reasons. Firstly, other than seen in the experiment, no superimposed undirected translational motion is considered. The model does not contain any inhomogenities regarding the particle geometry and heat absorption, which numerically results in perfect rotation of the microgears from the start. Secondly, we assume a linear light absorption, which does not take possible nonlinear optical effects for low illumination powers into consideration. Furthermore, we neglect light absorption by the liquid. This might lead to unconsidered initial threshold effects. However, the general rotation ranges are in accordance with the experimental data, which enables well-founded predictions. In addition, for relative comparisons of different properties or geometric configurations, the model is valid.
\section{\label{sec:rotmot}Characterization of rotational microgear motion\protect\\}
In the following section, the influence of geometry and particle size on the rotation of microgears is numerically determined. The parameterization as in table\,\ref{props} is used for all calculations.  Furthermore, a light absorption coefficient of 2\,\% is assumed.

\subsection{Influence of the number of teeth}
\begin{figure}[H]
    \centerline{\includegraphics[scale=0.8]{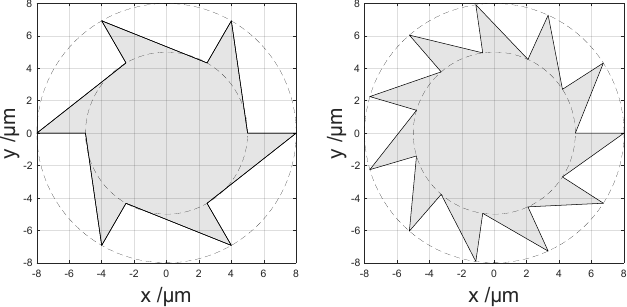}}
    \caption{Microgear geometries for a configuration with six (left) and eleven (right) teeth. The inner and outer radii are kept constant at 5\,µm and 8\,µm, respectively.}
    \label{fig:geomrot}
\end{figure}

\begin{figure}[H]
    \centerline{\includegraphics[scale=0.6]{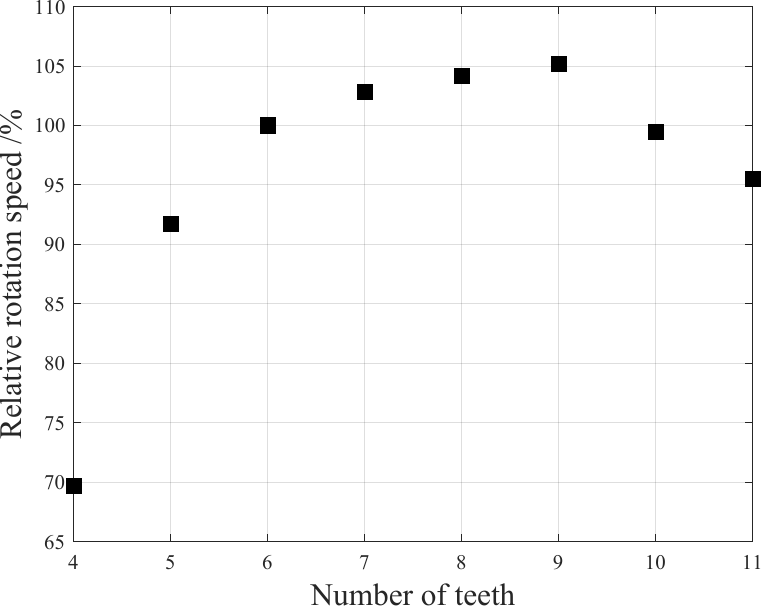}}
    \caption{Comparison of rotational speeds for a varying number of teeth, relative to the speed of a six-toothed microgear. For low numbers of teeth, rotational speed can be increased when adding a tooth.}
    \label{fig:teethnumber}
\end{figure}

Microgears with varying numbers of teeth have been analysed in existing literature\cite{maggi2015micromotors,yang2014self}. However, the influence of the geometric configuration has not been  examined in the literature thus far. To investigate the geometric influence, microgears with different numbers of teeth but constant inner and outer radii are compared. Figure\,\ref{fig:geomrot} illustrates different geometries for a microgear with six and eleven teeth.
\noindent
The absorbed illumination power is scaled in accordance with the surface area of the particles, as described in equation\,\eqref{arearatio}. This approach enables a systematic exploration of how variations in gear geometry impact rotational performance, facilitating the optimization of microgear design for improved efficiency.
Simulations show that for low numbers of teeth, the rotational speed of the microgear increases with teeth number, as seen in figure\,\ref{fig:teethnumber}. However, at some point, there is a limitation to this effect, and the rotational speed decreases again.

\begin{figure}[H]
    \centerline{\includegraphics[scale=0.6]{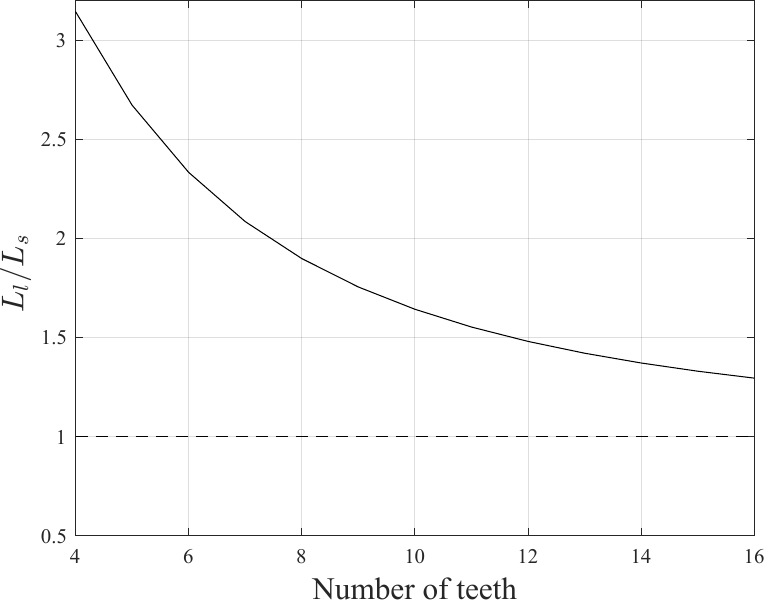}}
    \caption{When increasing the number of teeth, the length ratio of the long side $L_{l}$ and the short side $L_{s}$ approaches a value of one. This will lead to symmetric teeth and a net torque of zero.}
    \label{fig:ratio}
\end{figure}

\noindent
This occurrence of an optimal number of teeth can be explained by the following counteracting effects. The surface tension, defined as a force per unit length, implies that a larger circumference offers greater potential for applying forces. Hence, provided that the forces are always of the same order of magnitude, more teeth, i.e. a larger circumference, should lead to a higher velocity. However, the distribution of these forces is crucial. In symmetric configurations, forces tend to cancel each other out, regardless of their magnitude. Thus, besides circumference, the degree of asymmetry significantly influences rotational speed. These  factors counteract each other. While a higher number of teeth results in a longer circumference, the geometry approaches symmetry, diminishing rotational speed. As the number of teeth $n$ increases towards infinity, both sides of the tooth approach equal length (see figure\,\ref{fig:ratio}), leading to a symmetric shape of the teeth. Since the torques on both sides of each tooth are acting in opposite directions, the rotational speed tends towards zero.
\noindent
Moreover, additional factors impede rotation. The microgear's moment of inertia increases with each tooth, and greater surface area increases friction. Additionally, evaluation of the temperature along the edges of the teeth yields that the temperature reaches a saturation with added teeth. For a microgear configuration with more than nine teeth, these factors lead to a reduction of the total torque acting on the gear and therefore to a reduction of the thermocapillary rotation.

\subsection{Dimensional limitation of applicability}
\begin{figure}[H]
    \centerline{\includegraphics[scale=0.6]{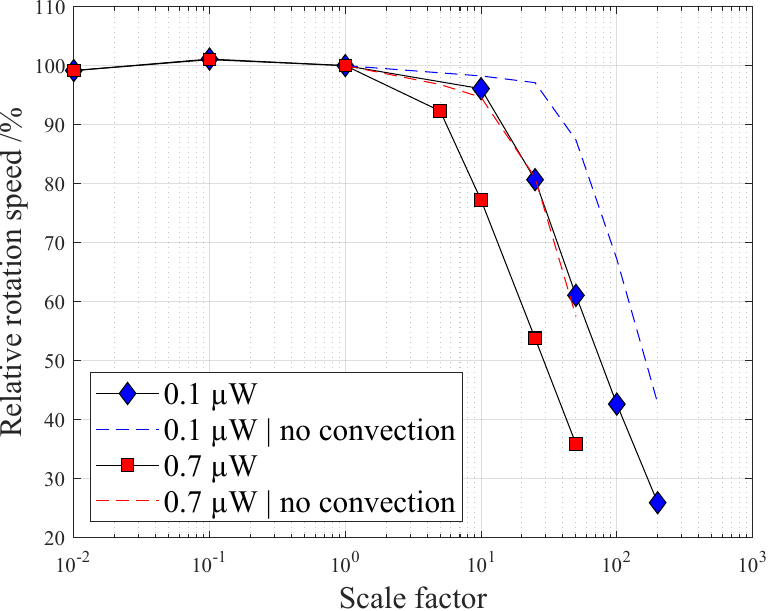}}
    \caption{Rotational speeds of a six-toothed microgear of different sizes for low and high illumination, relative to a 16\,µm large particle (scale factor 1). }
    \label{fig:dim}
\end{figure}

\begin{figure}[H]
    \centerline{\includegraphics[scale=0.6]{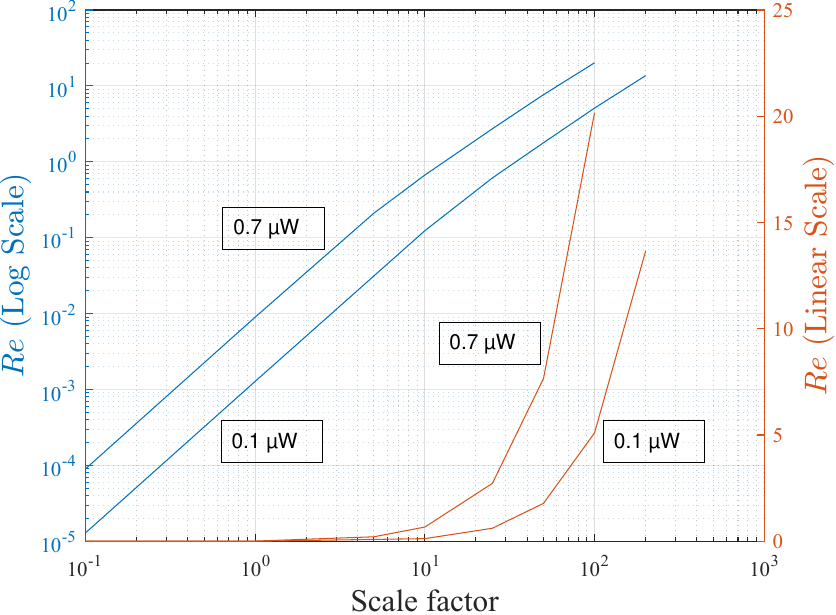}}
    \caption{Reynolds numbers for low and high illumination with the maximum velocity of the fluid domain as characteristic speed $U$ and the particle diameter as the characteristic length $L$.}
    \label{fig:Reynolds}
\end{figure}

The existence of thermocapillary rotation in asymmetric particles has been demonstrated on both nano- and microscales through various research studies\cite{maggi2015micromotors,liu2022light}. Maggi et al. propose that since system size is not explicitly incorporated into their derived estimation equation for angular speed, illumination-induced rotation might be achievable across a broader range of size scales\cite{maggi2015micromotors}. To explore the boundaries of this rotation, simulations are conducted across multiple size scales. The entire simulation domain is scaled using a scale factor (scf) ranging from 0.01 to 200. A scale factor of 1 refers to a particle diameter of 16\,µm, as utilized in the aforementioned studies. The particle's moment of inertia and absorbed illumination power are once again adjusted accordingly to maintain a constant incident illumination power across all size scales. This corresponds to an experimental scenario in which particles of different sizes are illuminated by the same light source. Confirming the assumption of Maggi et al., comparable rotational speeds are indeed achieved for smaller size scales, that is below the reference radius of 16\,µm.  The rotational speed remains relatively constant for scale factors of 0.1 and even 0.01 (see figure\,\ref{fig:dim}). However, it is questionable to which extent the continuum assumption, upon which the simulations are based, is violated for such small scaling factors. In practice, unconsidered nanoeffects might exert a superimposed influence on the motion.
\noindent
For larger particles than the reference size, the rotational speed significantly decreases. This can be explained by two reasons. First, the influence of heat convection in the liquid becomes prominent due to the presence of higher absolute temperature gradients on larger size scales. This convection results in a more distributed heat pattern within the fluid domain, consequently reducing the temperature gradients and thus the thermocapillary forces, which in turn decelerates particle motion. Simulations show that the relative average temperature gradients $\Delta T_{\mathrm{av}}/\mathrm{scf}=(T_{\mathrm{av}}-293.15\,\,\mathrm{K})/\mathrm{scf}$ between the three-phase contact line around the particle and the imposed ambient temperature at the fluid boundary remain constant for scale factors of one and below. For larger particles, convection leads to smaller gradients relative to the particle size, which initially accounts for the notable decline observed in the rotational speed (depicted by the drop in the blue- and red-marked black curves in figure\,\ref{fig:dim}). This effect is further evidenced by the comparison with simulations neglecting heat convection, setting $\rho c_{p} \vec{u} \nabla T=0$ for the fluid in equation\,\eqref{eq:heattransfer}. There, the drop in the curve is less steep, indicating a relatively constant rotational speed for larger size scales (as depicted by the dashed blue and red curves in figure\,\ref{fig:dim}). However, with a further increase in particle size, other effects become predominant in causing the decline in rotational speed. Specifically, the relative maximum velocity in the fluid domain $u_{max}/\mathrm{scf}$ decreases even more rapidly than the abovementioned relative temperature gradients, indicating a shift in the flow regime characterized by the Reynolds number $Re$. The Reynolds number relates inertial and viscous forces 
\begin{equation}
    Re=\frac{\rho U L} {\mu}.
\end{equation}
$U$ and $L$ represent a characteristic speed and a characteristic length, respectively. At low Reynolds numbers ${Re\ll1}$, viscous forces like Marangoni stresses dominate. However, as the Reynolds number increases, and inertial effects rise, their significance decreases.
When comparing the rotational speeds with the corresponding Reynolds numbers in figure\,\ref{fig:Reynolds}, this relationship is clearly visible. As soon as the Reynolds number exceeds unity (${Re>1}$), indicating a transition where inertial forces start to dominate, the rotational speed experiences a significant drop.

\noindent
Hence, the decline in rotational speed is more pronounced for higher illumination powers compared to lower illuminations (see blue- and red-marked black curves in figure\,\ref{fig:dim}). The transition from Stokes flow to a flow regime characterized by Reynolds numbers greater than one occurs at relatively smaller scale factors due to the higher absolute fluid velocity. Even relatively small microgears can cause a scenario with high enough Reynolds numbers to drop out of a Stokes regime, when rotating fast enough. This suggests that when utilizing thermocapillary forces for particle rotation on size scales above the micron scale, the efficiency diminishes with increasing illumination power. Additionally, on the micron scale, this implies a natural limitation to the linearity between rotational speed and illumination power.

\section{\label{sec:translational}Translational motion\protect\\}
\begin{figure}[H]
    \centerline{\includegraphics[scale=0.33]{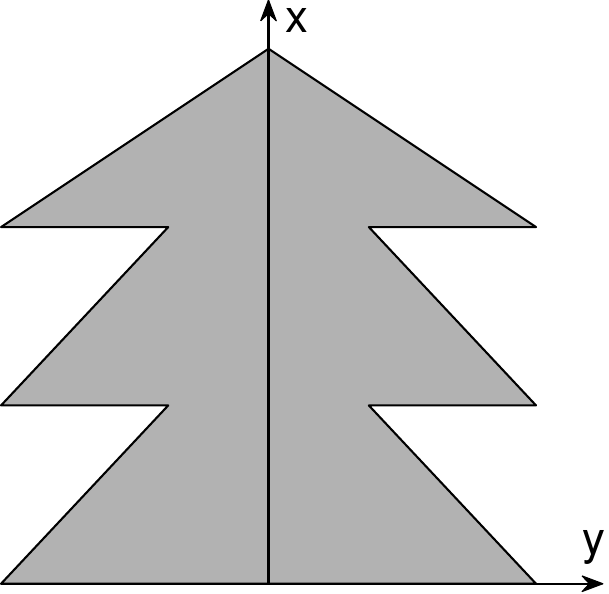}}
    \caption{2D view of a translational microswimmer with three spikes (per side). Symmetry is broken in the $x$-direction.}
    \label{fig:geom3}
\end{figure}

As noted in section \ref{sec:introduction}, existing literature primarily focuses on half-coated Janus particles for translational propulsion\cite{PhysRevLett.125.098001,masoud2014reciprocal}, achieving asymmetric surface tension distribution through coating. The basic concept behind the translational propulsion presented in this work is to leverage geometric asymmetry to induce forces in a direction of translation, rather than creating a net torque. For example, in the case of a 2D particle, a shape that is symmetric in $y$-direction but asymmetric in $x$-direction will, under similar illumination conditions as in the case of microgears, lead to a unidimensional imbalance of forces in $x$-direction, thus initiating motion in that direction. Based on a microgear’s tooth, a simple “Christmas-tree”-shaped particle (see figure\,\ref{fig:geom3}) fulfills these criteria. 
\noindent

\begin{figure}[H]
    \centerline{\includegraphics[scale=0.626]{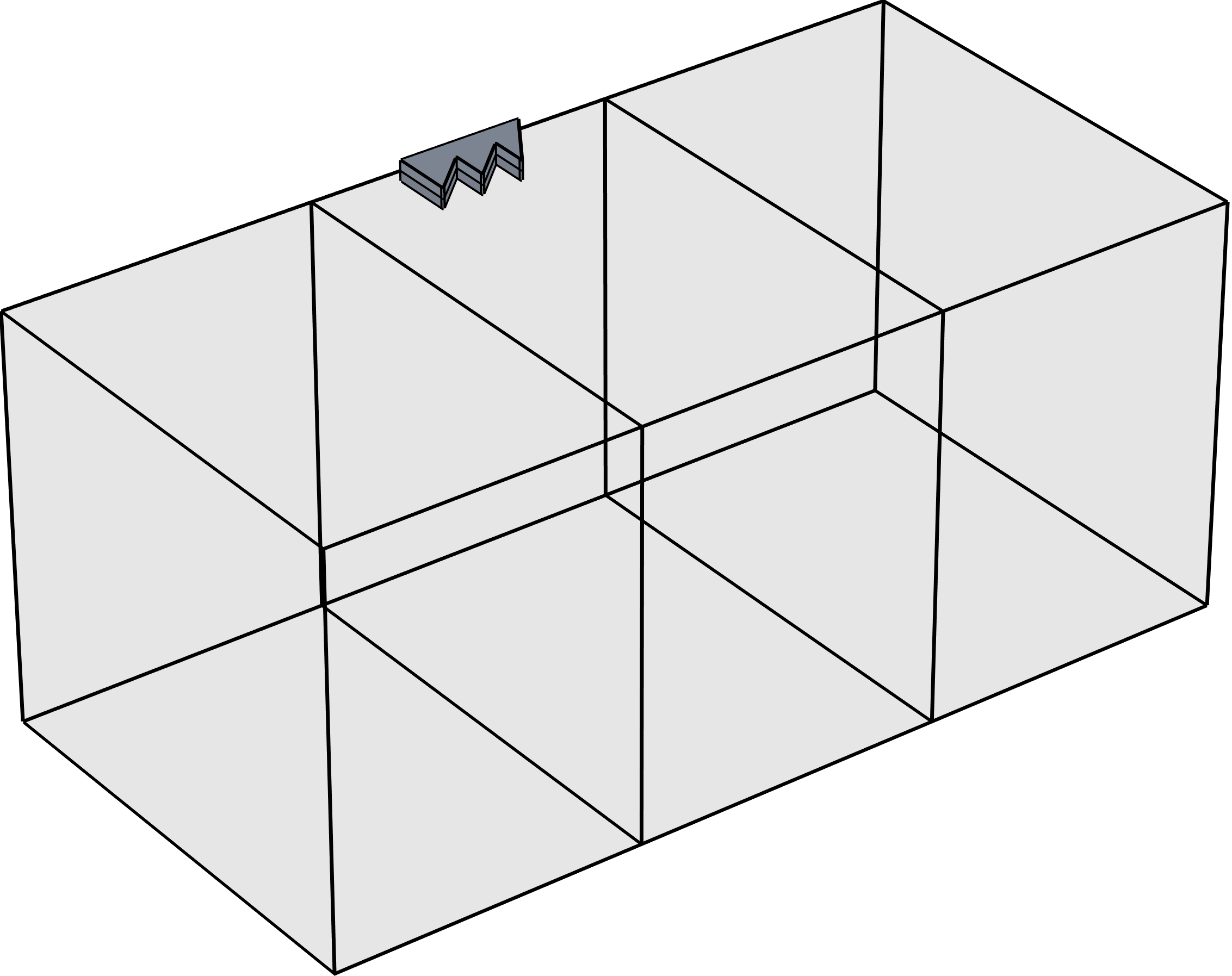}}
    \caption{Simulation domain for translational microswimmers. The mesh of the middle part of the domain will move equally with the particle to maintain a non-distorted surrounding mesh.}
    \label{fig:domaintrans}
\end{figure}

\begin{figure}[H]
    \centerline{\includegraphics[scale=0.6]{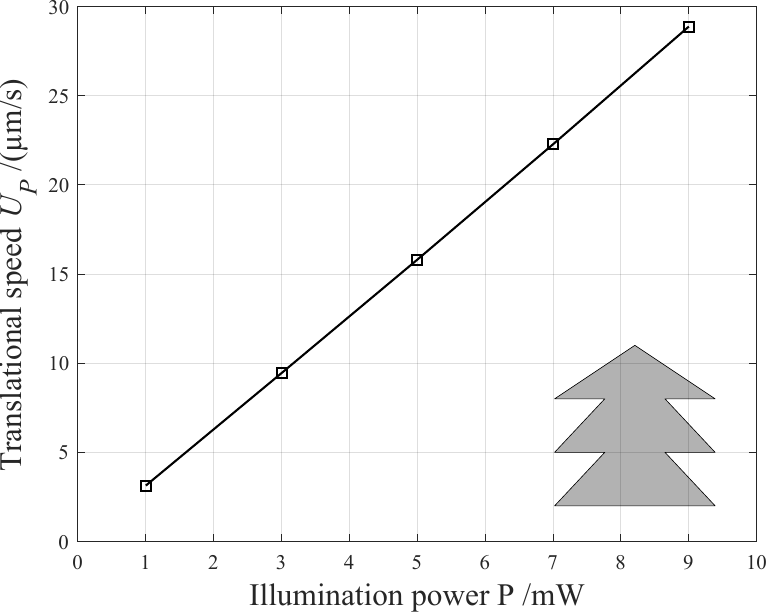}}
    \caption{As in case of thermocapillary rotation, linearity between translational propulsion speed and illumination power is recognizable. }
    \label{fig:translinspeed}
\end{figure}

\begin{figure}[H]
    \centerline{\includegraphics[scale=0.6]{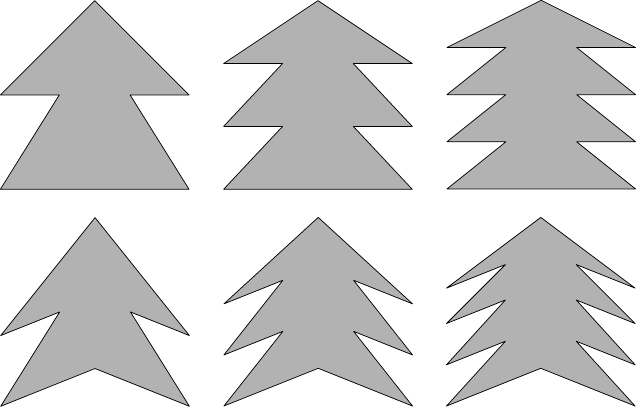}}
    \caption{Translational microswimmer geometries with two, three and four spikes, straight and inclined edges. The measurements are kept constant with 16 µm in length and in width. For the simulations, the edges are rounded with a radius of 0.1\,µm.}
    \label{fig:allgeom}
\end{figure}

For the most part, the modeling principles can be transferred from rotational to translational motion.
Instead of rotational dynamics, a translational velocity $U_{P}$ is calculated based on the forces acting in $x$-direction $F_{x,\mathrm{tot}}$. Mathematically, this leads to a replacement of equations \eqref{dynamicrot} and \eqref{mom} with 
 \begin{equation}
	\frac{\partial U_{P,x}}{\partial t}=\frac{F_{x,\mathrm{tot}}}{m},
	\label{dynamictrans}
\end{equation}
and
\begin{equation}
F_{x,\mathrm{tot}}=\int_{A_{c}} p\cdot n_{x} \,\mathrm{d}A_{c} - \int_{A_{c}} \tau_{x} \,\mathrm{d}A_{c},
    \label{xforce}
\end{equation}
with the first term of equation\,\eqref{xforce} being the normal pressure force and the second term representing the viscous force, acting in $x$-direction on the particle wall. Here, $m$ represents the mass of the particle and $n_{x}$ is the $x$-component of the normal vector on the particle surface. Furthermore, the simulation domain changes (see figure\,\ref{fig:domaintrans}). To save simulation time, symmetry can be exploited in case of translation.
\noindent
The simulations show that, when illuminating the translational microswimmer, the thermocapillary forces lead to a forward motion of the particle. As in the case of rotation, the propulsion speed increases linearly with the illumination power, as depicted in figure\,\ref{fig:translinspeed}.
\noindent
 Analogous to a microgear's number of teeth, the speed is expected to be influenced by the number of spikes. To quantify the influence, microswimmers with different numbers of spikes (see figure\,\ref{fig:allgeom}) are examined. Furthermore, an inclination of the spike edges, which is assumed to accelerate particle motion, is compared with the straight-edged geometries.

\begin{figure}[H]
    \centerline{\includegraphics[scale=0.6]{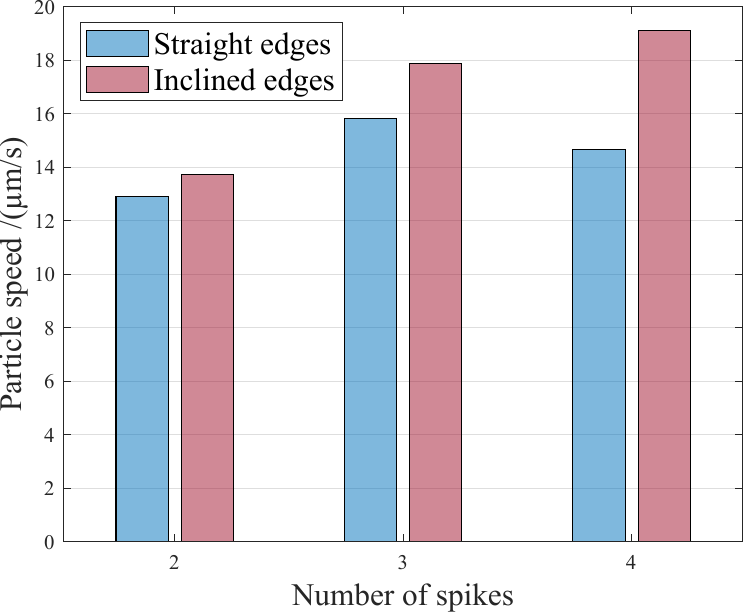}}
    \caption{Particle propulsion speeds of various geometric configurations, referring to an illumination power P of 5\,mW.}
    \label{fig:transspeed}
\end{figure}

\noindent
As expected, all of the particle geometries lead to a forward motion. The influence of the number of spikes is recognizable and of a comparable order of magnitude to the influence of the number of teeth in a microgear rotation scenario. Interestingly, a inclination of the microswimmer's spikes seems to accelerate motion regardless of the number of spikes.  Figure\,\ref{fig:transspeed} indicates that the effect of spike inclination on the propulsion speed increases with the number of spikes. We assume that this effect becomes even stronger when further inclining the spikes.

\noindent
Comparing the results to a rotating microgear scenario as in section \ref{sec:ValRes}, the speeds in figure\,\ref{fig:transspeed} at 2\,mW correspond to a rotational speed of around 14\,rpm. Considering the particle measurements of 16\,µm, due to the linearity between speed and illumination, maximum particle speeds of several particle lengths per second will be possible. Even a propulsion speed of ten times the particle length per second might be achievable in an experimental setup.

\section{\label{sec:conclusion}Conclusion\protect\\}
In the first part of our work, we have numerically validated previous experimental findings regarding illumination-induced rotation of micron-sized particles. The rotation rates predicted align closely with those observed in the experimental work by Maggi et al.\cite{maggi2015micromotors}. We observe that increasing the gear’s number of teeth can lead to a decrease in rotational speed, despite the larger circumference, primarily due to approaching symmetry in the gear's geometry.
Regarding size scale, our findings indicate that on the lower micro- and nanoscale, the size of the gears does not significantly affect rotational speed. However, for size scales in the higher micrometer range and beyond, there is a notable decrease in rotational speed. This phenomenon can be attributed to heat convection in the liquid and the transition to a flow regime characterized by a Reynolds number greater than one, where viscous forces become less dominant.

Moreover, we have presented the transferability of geometric symmetry-breaking to translational thermocapillary particle motion. Depending on the particle design and illumination power, our simulations demonstrate that these translational swimmers can achieve propulsion speeds of several particle lengths per second. This highlights the potential of leveraging geometric asymmetry for efficient directed light-driven translational motion, paving the way for diverse applications in micro- and nanotechnology.

\begin{acknowledgments}
We kindly acknowledge the funding by the Deutsche Forschungsgemeinschaft (DFG, German Research Foundation) - project number 501107071.
\end{acknowledgments}

\nocite{*}
\bibliography{aipsamp}

\begin{thebibliography}{11}%
\makeatletter
\providecommand \@ifxundefined [1]{%
 \@ifx{#1\undefined}
}%
\providecommand \@ifnum [1]{%
 \ifnum #1\expandafter \@firstoftwo
 \else \expandafter \@secondoftwo
 \fi
}%
\providecommand \@ifx [1]{%
 \ifx #1\expandafter \@firstoftwo
 \else \expandafter \@secondoftwo
 \fi
}%
\providecommand \natexlab [1]{#1}%
\providecommand \enquote  [1]{``#1''}%
\providecommand \bibnamefont  [1]{#1}%
\providecommand \bibfnamefont [1]{#1}%
\providecommand \citenamefont [1]{#1}%
\providecommand \href@noop [0]{\@secondoftwo}%
\providecommand \href [0]{\begingroup \@sanitize@url \@href}%
\providecommand \@href[1]{\@@startlink{#1}\@@href}%
\providecommand \@@href[1]{\endgroup#1\@@endlink}%
\providecommand \@sanitize@url [0]{\catcode `\\12\catcode `\$12\catcode `\&12\catcode `\#12\catcode `\^12\catcode `\_12\catcode `\%12\relax}%
\providecommand \@@startlink[1]{}%
\providecommand \@@endlink[0]{}%
\providecommand \url  [0]{\begingroup\@sanitize@url \@url }%
\providecommand \@url [1]{\endgroup\@href {#1}{\urlprefix }}%
\providecommand \urlprefix  [0]{URL }%
\providecommand \Eprint [0]{\href }%
\providecommand \doibase [0]{http://dx.doi.org/}%
\providecommand \selectlanguage [0]{\@gobble}%
\providecommand \bibinfo  [0]{\@secondoftwo}%
\providecommand \bibfield  [0]{\@secondoftwo}%
\providecommand \translation [1]{[#1]}%
\providecommand \BibitemOpen [0]{}%
\providecommand \bibitemStop [0]{}%
\providecommand \bibitemNoStop [0]{.\EOS\space}%
\providecommand \EOS [0]{\spacefactor3000\relax}%
\providecommand \BibitemShut  [1]{\csname bibitem#1\endcsname}%
\let\auto@bib@innerbib\@empty
\bibitem [{\citenamefont {Maggi}\ \emph {et~al.}(2015)\citenamefont {Maggi}, \citenamefont {Saglimbeni}, \citenamefont {Dipalo}, \citenamefont {De~Angelis},\ and\ \citenamefont {Di~Leonardo}}]{maggi2015micromotors}%
  \BibitemOpen
  \bibfield  {author} {\bibinfo {author} {\bibfnamefont {C.}~\bibnamefont {Maggi}}, \bibinfo {author} {\bibfnamefont {F.}~\bibnamefont {Saglimbeni}}, \bibinfo {author} {\bibfnamefont {M.}~\bibnamefont {Dipalo}}, \bibinfo {author} {\bibfnamefont {F.}~\bibnamefont {De~Angelis}}, \ and\ \bibinfo {author} {\bibfnamefont {R.}~\bibnamefont {Di~Leonardo}},\ }\bibfield  {title} {\enquote {\bibinfo {title} {Micromotors with asymmetric shape that efficiently convert light into work by thermocapillary effects},}\ }\href@noop {} {\bibfield  {journal} {\bibinfo  {journal} {Nature communications}\ }\textbf {\bibinfo {volume} {6}},\ \bibinfo {pages} {7855} (\bibinfo {year} {2015})}\BibitemShut {NoStop}%
\bibitem [{\citenamefont {S{\'a}nchez}, \citenamefont {Soler},\ and\ \citenamefont {Katuri}(2015)}]{sanchez2015chemically}%
  \BibitemOpen
  \bibfield  {author} {\bibinfo {author} {\bibfnamefont {S.}~\bibnamefont {S{\'a}nchez}}, \bibinfo {author} {\bibfnamefont {L.}~\bibnamefont {Soler}}, \ and\ \bibinfo {author} {\bibfnamefont {J.}~\bibnamefont {Katuri}},\ }\bibfield  {title} {\enquote {\bibinfo {title} {Chemically powered micro-and nanomotors},}\ }\href@noop {} {\bibfield  {journal} {\bibinfo  {journal} {Angewandte Chemie International Edition}\ }\textbf {\bibinfo {volume} {54}},\ \bibinfo {pages} {1414--1444} (\bibinfo {year} {2015})}\BibitemShut {NoStop}%
\bibitem [{\citenamefont {Wang}\ \emph {et~al.}(2013)\citenamefont {Wang}, \citenamefont {Duan}, \citenamefont {Ahmed}, \citenamefont {Mallouk},\ and\ \citenamefont {Sen}}]{wang2013small}%
  \BibitemOpen
  \bibfield  {author} {\bibinfo {author} {\bibfnamefont {W.}~\bibnamefont {Wang}}, \bibinfo {author} {\bibfnamefont {W.}~\bibnamefont {Duan}}, \bibinfo {author} {\bibfnamefont {S.}~\bibnamefont {Ahmed}}, \bibinfo {author} {\bibfnamefont {T.~E.}\ \bibnamefont {Mallouk}}, \ and\ \bibinfo {author} {\bibfnamefont {A.}~\bibnamefont {Sen}},\ }\bibfield  {title} {\enquote {\bibinfo {title} {Small power: Autonomous nano-and micromotors propelled by self-generated gradients},}\ }\href@noop {} {\bibfield  {journal} {\bibinfo  {journal} {Nano Today}\ }\textbf {\bibinfo {volume} {8}},\ \bibinfo {pages} {531--554} (\bibinfo {year} {2013})}\BibitemShut {NoStop}%
\bibitem [{\citenamefont {Yang}\ and\ \citenamefont {Ripoll}(2014)}]{yang2014self}%
  \BibitemOpen
  \bibfield  {author} {\bibinfo {author} {\bibfnamefont {M.}~\bibnamefont {Yang}}\ and\ \bibinfo {author} {\bibfnamefont {M.}~\bibnamefont {Ripoll}},\ }\bibfield  {title} {\enquote {\bibinfo {title} {A self-propelled thermophoretic microgear},}\ }\href@noop {} {\bibfield  {journal} {\bibinfo  {journal} {Soft Matter}\ }\textbf {\bibinfo {volume} {10}},\ \bibinfo {pages} {1006--1011} (\bibinfo {year} {2014})}\BibitemShut {NoStop}%
\bibitem [{\citenamefont {Angelani}, \citenamefont {Di~Leonardo},\ and\ \citenamefont {Ruocco}(2009)}]{angelani2009self}%
  \BibitemOpen
  \bibfield  {author} {\bibinfo {author} {\bibfnamefont {L.}~\bibnamefont {Angelani}}, \bibinfo {author} {\bibfnamefont {R.}~\bibnamefont {Di~Leonardo}}, \ and\ \bibinfo {author} {\bibfnamefont {G.}~\bibnamefont {Ruocco}},\ }\bibfield  {title} {\enquote {\bibinfo {title} {Self-starting micromotors in a bacterial bath},}\ }\href@noop {} {\bibfield  {journal} {\bibinfo  {journal} {Physical review letters}\ }\textbf {\bibinfo {volume} {102}},\ \bibinfo {pages} {048104} (\bibinfo {year} {2009})}\BibitemShut {NoStop}%
\bibitem [{\citenamefont {Dietrich}\ \emph {et~al.}(2020)\citenamefont {Dietrich}, \citenamefont {Jaensson}, \citenamefont {Buttinoni}, \citenamefont {Volpe},\ and\ \citenamefont {Isa}}]{PhysRevLett.125.098001}%
  \BibitemOpen
  \bibfield  {author} {\bibinfo {author} {\bibfnamefont {K.}~\bibnamefont {Dietrich}}, \bibinfo {author} {\bibfnamefont {N.}~\bibnamefont {Jaensson}}, \bibinfo {author} {\bibfnamefont {I.}~\bibnamefont {Buttinoni}}, \bibinfo {author} {\bibfnamefont {G.}~\bibnamefont {Volpe}}, \ and\ \bibinfo {author} {\bibfnamefont {L.}~\bibnamefont {Isa}},\ }\bibfield  {title} {\enquote {\bibinfo {title} {Microscale marangoni surfers},}\ }\href {\doibase 10.1103/PhysRevLett.125.098001} {\bibfield  {journal} {\bibinfo  {journal} {Phys. Rev. Lett.}\ }\textbf {\bibinfo {volume} {125}},\ \bibinfo {pages} {098001} (\bibinfo {year} {2020})}\BibitemShut {NoStop}%
\bibitem [{\citenamefont {Masoud}\ and\ \citenamefont {Stone}(2014)}]{masoud2014reciprocal}%
  \BibitemOpen
  \bibfield  {author} {\bibinfo {author} {\bibfnamefont {H.}~\bibnamefont {Masoud}}\ and\ \bibinfo {author} {\bibfnamefont {H.~A.}\ \bibnamefont {Stone}},\ }\bibfield  {title} {\enquote {\bibinfo {title} {A reciprocal theorem for marangoni propulsion},}\ }\href@noop {} {\bibfield  {journal} {\bibinfo  {journal} {Journal of Fluid Mechanics}\ }\textbf {\bibinfo {volume} {741}},\ \bibinfo {pages} {R4} (\bibinfo {year} {2014})}\BibitemShut {NoStop}%
\bibitem [{\citenamefont {Sur}, \citenamefont {Masoud},\ and\ \citenamefont {Rothstein}(2019)}]{sur2019translational}%
  \BibitemOpen
  \bibfield  {author} {\bibinfo {author} {\bibfnamefont {S.}~\bibnamefont {Sur}}, \bibinfo {author} {\bibfnamefont {H.}~\bibnamefont {Masoud}}, \ and\ \bibinfo {author} {\bibfnamefont {J.~P.}\ \bibnamefont {Rothstein}},\ }\bibfield  {title} {\enquote {\bibinfo {title} {Translational and rotational motion of disk-shaped marangoni surfers},}\ }\href@noop {} {\bibfield  {journal} {\bibinfo  {journal} {Physics of Fluids}\ }\textbf {\bibinfo {volume} {31}} (\bibinfo {year} {2019})}\BibitemShut {NoStop}%
\bibitem [{\citenamefont {Crowdy}(2021)}]{crowdy2021viscous}%
  \BibitemOpen
  \bibfield  {author} {\bibinfo {author} {\bibfnamefont {D.}~\bibnamefont {Crowdy}},\ }\bibfield  {title} {\enquote {\bibinfo {title} {Viscous propulsion of a two-dimensional marangoni boat driven by reaction and diffusion of insoluble surfactant},}\ }\href@noop {} {\bibfield  {journal} {\bibinfo  {journal} {Physical Review Fluids}\ }\textbf {\bibinfo {volume} {6}},\ \bibinfo {pages} {064003} (\bibinfo {year} {2021})}\BibitemShut {NoStop}%
\bibitem [{\citenamefont {Kang}\ \emph {et~al.}(2020)\citenamefont {Kang}, \citenamefont {Sur}, \citenamefont {Rothstein},\ and\ \citenamefont {Masoud}}]{kang2020forward}%
  \BibitemOpen
  \bibfield  {author} {\bibinfo {author} {\bibfnamefont {S.~J.}\ \bibnamefont {Kang}}, \bibinfo {author} {\bibfnamefont {S.}~\bibnamefont {Sur}}, \bibinfo {author} {\bibfnamefont {J.~P.}\ \bibnamefont {Rothstein}}, \ and\ \bibinfo {author} {\bibfnamefont {H.}~\bibnamefont {Masoud}},\ }\bibfield  {title} {\enquote {\bibinfo {title} {Forward, reverse, and no motion of marangoni surfers under confinement},}\ }\href@noop {} {\bibfield  {journal} {\bibinfo  {journal} {Physical Review Fluids}\ }\textbf {\bibinfo {volume} {5}},\ \bibinfo {pages} {084004} (\bibinfo {year} {2020})}\BibitemShut {NoStop}%
\bibitem [{\citenamefont {Liu}\ \emph {et~al.}(2022)\citenamefont {Liu}, \citenamefont {Jiang}, \citenamefont {Zhu}, \citenamefont {Li}, \citenamefont {Zhang}, \citenamefont {Tian}, \citenamefont {Wang}, \citenamefont {Wang}, \citenamefont {Ouyang}, \citenamefont {Xiao} \emph {et~al.}}]{liu2022light}%
  \BibitemOpen
  \bibfield  {author} {\bibinfo {author} {\bibfnamefont {C.}~\bibnamefont {Liu}}, \bibinfo {author} {\bibfnamefont {D.}~\bibnamefont {Jiang}}, \bibinfo {author} {\bibfnamefont {G.}~\bibnamefont {Zhu}}, \bibinfo {author} {\bibfnamefont {Z.}~\bibnamefont {Li}}, \bibinfo {author} {\bibfnamefont {X.}~\bibnamefont {Zhang}}, \bibinfo {author} {\bibfnamefont {P.}~\bibnamefont {Tian}}, \bibinfo {author} {\bibfnamefont {D.}~\bibnamefont {Wang}}, \bibinfo {author} {\bibfnamefont {E.}~\bibnamefont {Wang}}, \bibinfo {author} {\bibfnamefont {H.}~\bibnamefont {Ouyang}}, \bibinfo {author} {\bibfnamefont {M.}~\bibnamefont {Xiao}},  \emph {et~al.},\ }\bibfield  {title} {\enquote {\bibinfo {title} {A light-powered triboelectric nanogenerator based on the photothermal marangoni effect},}\ }\href@noop {} {\bibfield  {journal} {\bibinfo  {journal} {ACS Applied Materials \& Interfaces}\ }\textbf {\bibinfo {volume} {14}},\ \bibinfo {pages} {22206--22215} (\bibinfo {year} {2022})}\BibitemShut {NoStop}%
\end{thebibliography}%

\end{document}